# Structuring Authenticity Assessments on Historical Documents using LLMs

Andrea Schimmenti[2], Valentina Pasqual[1], Francesca Tomasi[1], Fabio Vitali[2] and Marieke van Erp[3]

[1] Digital Humanities Advanced Research Center, (/DH.arc), FICLIT, University of Bologna, Italy - valentina.pasqual2@unibo.it, francesca.tomasi@unibo.it

[2] University of Bologna, Italy - andrea.schimmenti2@unibo.it, fabio.vitali@unibo.it

[3] DHLab - KNAW Humanities Cluster, Netherlands - marieke.van.erp@dh.huc.knaw.nl

**ABSTRACT**

Given the wide use of forgery throughout history, scholars have and are continuously engaged in assessing the authenticity of historical documents. However, online catalogues merely offer descriptive metadata for these documents, relegating discussions about their authenticity to free-text formats, making it difficult to study these assessments at scale. This study explores the generation of structured data about documents' authenticity assessment from natural language texts. Our pipeline exploits Large Language Models (LLMs) to select, extract and classify relevant claims about the topic without the need for training, and Semantic Web technologies to structure and type-validate the LLM's results. The final output is a catalogue of documents whose authenticity has been debated, along with scholars' opinions on their authenticity. This process can serve as a valuable resource for integration into catalogues, allowing room for more intricate queries and analyses on the evolution of these debates over centuries.

## 1. INTRODUCTION

Historical authenticity assessment is a scholarly practice used to determine the authenticity of historical documents. Scholars from different humanities and scientific disciplines (e.g. Diplomatics, Palaeography, Philology, History, Forensics) have contributed to the task [2]. Various scholars arrive at divergent and possibly contrasting conclusions due to considering different evidence. Inherent factors contributing to this diversity include historical uncertainty, gaps in documentary transmission, and subjectivity [3,9]. For example, the *Donation of Constantine* has been studied by several scholars from different perspectives, and its authenticity has been widely discussed for centuries.[1] The debate revolves around this being a supposed decree by Constantine from the 4th century until scholar Lorenzo Valla claimed it was a later forgery at the beginning of the Renaissance. To the best of our knowledge, such complex information is not included in digital libraries, catalogues and archives. Considering the provided guiding example, the document can be found in Wikidata and DBpedia, but no mention of the debate is available in a structured format. Considering the intricate language and data sparseness of humanities discourse, the recent spread of Large Language Models (LLMs) widens the possibility of retrieval from unstructured natural language descriptions. Many tests have proved that LLMs perform promisingly when dealing with structured data extraction from unstructured text [13,14,16]. This work tackles the following research question: can LLMs be useful inside a pipeline as tools for extracting and classifying authenticity assessment debates, with the ultimate goal of structuring such claims into a KG? Our approach to answering this question revolves around analysing the authenticity assessments of Wikipedia articles using two different LLMs. We used the outputs of the LLMs to generate a KG. For the Donation of Constantine case study, we show how a structured authenticity assessment KG can automatically map a forgery debate.

## 2. RELATED WORK

Knowledge Graph (KG) generation from text has been the aim of many researchers inside and outside the Semantic Web community. For example, REBEL, trained on multiple languages (using a set of Wikidata properties and classes as ontology), is an example of state-of-the-art performance in generating simple KGs [12]. Since 2017, many NLP tasks have seen general improvements after introducing the Transformers architecture and pre-trained models such as BERT [8]. With their massive training on multiple corpora, Large Language Models can be useful, especially in cases where large datasets are unavailable for downstream tasks, e.g. KG generation. LLMs have also been tested for KG extraction from unstructured text, especially for KG generation and enrichment. BERT proved to have good generalisation capabilities, even with a few examples in the training data using in-context learning (i.e. describing a task alongside one or multiple examples) [14]. LLMs, such as the GPT family, have been tested for multiple tasks, achieving state of art results with Zero or Few-Shot learning [4]. However, LLMs are not completely reliable as a source of factual knowledge for sturdy KB enrichment, relying mostly on superficial natural language information. Yet, they perform well compared to state-of-the-art results in many NLP tasks [16]. Text2KG, a benchmark for evaluating LLM results over generating KGs given a simplified ontology and a few examples, has shown promising results when evaluated on two different datasets, Wikidata-TekGen and DBpedia-WebNLG [15]. The prompts used a simplified version of the Wikidata or the DBpedia set of properties, where a property was described as, e.g. "creatorOf(person, book)", a set of terms for the possible classes and a set of examples [15]. This type of in-context learning also shows interesting results when dealing with KG generation using GPT-3 [14]. In general, the application of LLMs in Knowledge Engineering is being discussed by the community with multiple open questions on how they can and should be used and how reliable they can be when dealing with knowledge engineering in general [1].

---
[1] https://en.wikipedia.org/wiki/Donation_of_Constantine

### 3. APPROACH

From a selection of Wikipedia articles regarding documentary forgery and medieval charters,[2] we extracted the document description and those sections tackling authenticity assessment. The test involved three tasks whose output was promptly evaluated. The output data created a KG using *forgont* as an ontology. For these tasks, we chose to test two different LLMs: GPT-4[3] and Llama version 2 70b[4]. The workflow of the experiment is as follows:

1. **Data modelling**. An ontology, called *forgont,* was defined based on a selection of scholarly articles [6,11], a catalogue describing 153 known forgeries from Styria [10], and collaboration with an expert Diplomatist. The ontology *forgont* represents authenticity assessment claims using the Named Graphs [5] structure as a reification method to represent both each (possibly concurrent) claim content and its contextual information [7]. The content of the claim addresses the following basic information about the document in the scholar's opinion: document authenticity classification (e.g. authentic, forgery, suspicious forgery), date and place of creation, author, and reason of the document. Additionally, contextual information about the claim is recorded, specifically the author of the claim, the document features observed by the scholar (e.g. ink, paper, style), the evidence found (e.g. inconsistent contents with other documents, lack of genuine witnesses) to reach a certain conclusion (e.g. the document is a forgery), and relevant sources.[5] The *forgont* structure served as a schema to define the prompts;
2. **LLM-assisted data extraction**.
    a. **Task 1: Document metadata extraction**. The LLM prompt asked to identify and extract a distinct set of properties from the text, namely the "document title", "document type", (alleged) "creation date", (alleged) "creation place," and (alleged) "author". The model is required to extract the metadata as presented *within* the forged document: in the case of the *Donation of Constantine*, the document was allegedly created during Constantine's reign, while it was in reality produced much later;
    b. **Task 2: Named Entity Recognition and Claim Identification**. The model is tasked to extract from the text every entity that is associated with a claim that discusses the document's authenticity;
    c. **Task 3: Claim classification**. Given the entity and the claim from the previous task, the model is instructed to classify the claims into three possible categories: "Authentic", "Forgery", and "Suspicious". The output is an object containing the author, class, opinion and source;

---

[2] https://en.wikipedia.org/wiki/Category:Document_forgery,
https://en.wikipedia.org/wiki/Category:Medieval_charters_and_cartularies and their subcategories
[3] https://openai.com/research/gpt-4
[4] https://huggingface.co/meta-llama/Llama-2-70b
[5] The complete documentation of both *forgont* model and relative Knowledge Base can be found at:
https://github.com/ValentinaPasqual/forgont/

3. **Evaluation.** LLM's performance for the metadata extraction and claim classification is measured using common metrics: precision, recall and $F_1$-score, while the claim identification is evaluated by the raw percentage of success. The number of ground truths was manually calculated, corresponding to 296 claims. The evaluation compares the two selected LLMs;
4. **Contents structuring and Knowledge Graph generation.** The contents extracted and classified by the winning LLM Data have been saved in JSON format and converted into RDF via Python scripts[6]. In the process, data have been cleaned and type-validated. In particular, people were semiautomatically reconciled against authority files (e.g. Viaf and Wikidata), dates were converted into machine-readable format, and places were reconciled and validated against Geonames. A human qualitative check of the dataset was performed to ensure quality. All claims were converted into RDF following the *forgont* schema.

## 4. EVALUATION

**First task: metadata extraction.** The data extraction task was evaluated on a corpus of 57 entries, assessing five key metadata properties: Title, Type, Date, Place, and Creator. Each entry of the corpus[7] contains the internal ID of the document, its Wikipedia link, its textual description and a selection of the sections describing the authenticity assessment debate. The criteria alignment is the following:
- Positive: The value was present and correctly identified in the text.
- Negative: The value was absent and correctly reported as not mentioned in the text.
- False Positive: An incorrect value is reported as present despite being absent in the text.
- False Negative: The value was present in the text but failed to be identified.

We computed Precision, Recall and F-1 score (Table 1).

| First task | Title | | Type | | Date | | Place | | Creator | |
|---|---|---|---|---|---|---|---|---|---|---|
| | GPT-4 | Llama-2 | GPT-4 | Llama-2 | GPT-4 | Llama-2 | GPT-4 | Llama-2 | GPT-4 | Llama-2 |
| **Precision** | 0.98 | 0.98 | 0.94 | 0.96 | 0.89 | 0.78 | 0.95 | 0.78 | 0.79 | 0.80 |
| **Recall** | 0.99 | 0.99 | 0.99 | 0.99 | 0.93 | 0.99 | 0.92 | 0.99 | 0.95 | 0.99 |
| $F_1$-score | 0.99 | 0.99 | 0.97 | 0.98 | 0.91 | 0.87 | 0.95 | 0.88 | 0.87 | 0.89 |

Table 1: Precision, Recall, $F_1$ (Metadata Extraction)

**Titles** and **Types** were mostly correctly identified and reported. The **Date** extraction performance implies some challenges in accurately extracting alleged dates, mistaking it for the forgery date. The models' performance vastly differed when identifying the alleged **Place** of origin or reference within the documents. The **Creator's** value was, as the date, sometimes confused with the author of the forgery. This last value gives much space for improvement. A drastic difference in performance is noticeable when authorship was not explicitly mentioned or was ambiguously referenced.

**Second task: Named Entity and Claim Identification.** This evaluation assessed if a model correctly recognised a named entity claiming something about a document's authenticity. An output is considered faulty if the entity was wrongly identified and/or if the text related to it was not a real claim, as shown in Table 2.

---

[6] All scripts are stored in the Github folder of the project.
[7] The selection is available in Github folder of the project.

|  | Correctly identified claims | Faulty or unidentified claims (%) |
|---|---|---|
| **GPT-4** | 254 | 42 (14.1%) |
| **Llama-2** | 226 | 70 (23.6%) |

Table 2: NER and claim extraction evaluation

**Third Task: Claim Categorisation Accuracy.** The third task involved a manual quantitative evaluation of the remaining claims (excluding faulty ones). This evaluation aimed to verify the correctness of the claim categorisation. To assess the model's performance in categorising claims as "Authentic", "Forgery", or "Suspicious", metrics such as F1-score, precision, and recall were utilised and computed based on a multiclass confusion matrix (Table 3). These metrics indicate a high level of accuracy in the model's ability to categorise claims, with good performance in identifying "Forgery" and "Suspicious" claims.

| Third task | Authentic | | Forgery | | Suspicious | |
|---|---|---|---|---|---|---|
|  | GPT-4 | Llama-2 | GPT-4 | Llama-2 | GPT-4 | Llama-2 |
| **Precision** | 0.87 | 0.74 | 0.97 | 0.92 | 0.94 | 0.88 |
| **Recall** | 0.83 | 0.86 | 0.90 | 0.80 | 0.96 | 0.88 |
| $F_1$-score | 0.85 | 0.80 | 0.94 | 0.85 | 0.95 | 0.88 |

Table 3: Precision, Recall, $F_1$ (Claim Categorisation Accuracy)

## 5. RESULT

The resulting KG stores 233 claims discussing the authenticity of 57 documents by 223 authors. Figure 1 exemplifies two claims (out of the 13 extracted) regarding the *Donation of Constantine*. Each claim is modelled as a Named Graph. On the one hand, the document claim - the result of extraction task 1 - contains the alleged author (emperor Constantine), the alleged date of creation (4th century, 300-399) and the document type (charter). On the other hand, the result of tasks 2 and 3 is shown in Valla's claim, which categorises the document as a forgery.

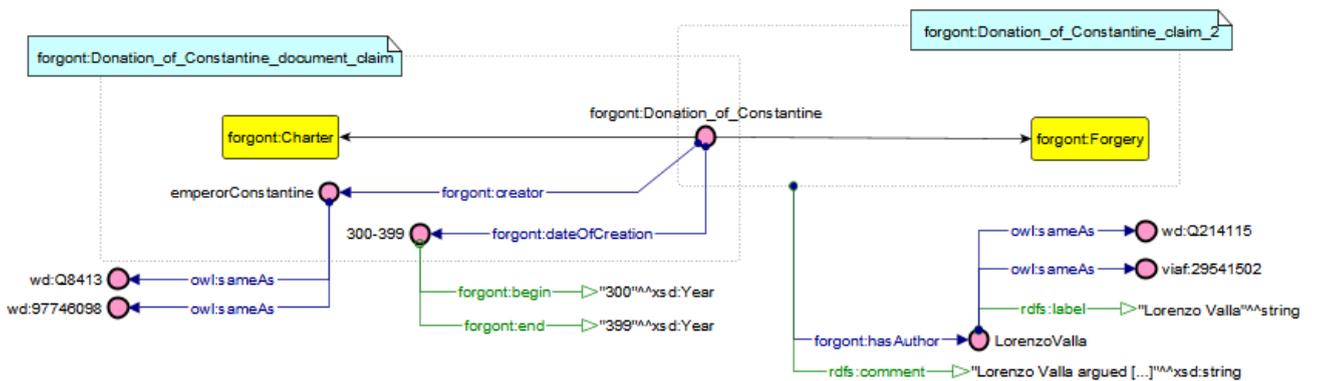

Figure 1: Example of extracted and structured claims regarding the *Donation of Constantine*

Organising, cleaning, and reconciling data facilitates data retrieval through SPARQL queries and establishes a foundation for in-depth data analysis and visualisation. Extracted claimers have been reconciled to authority files such as VIAF and Wikidata. Life dates of authors have been extracted from Wikidata, and the century of activity has been inferred. Figure 2 illustrates the case of the *Donation of Constantine* and all the claims extracted categories from Wikipedia and categorised by each claimer's century of activity (x-axis) with their opinion (colour). The figure constitutes a possible view of the data extracted and classified by this experiment for this document summarising the debate towards this charter authenticity over the centuries.

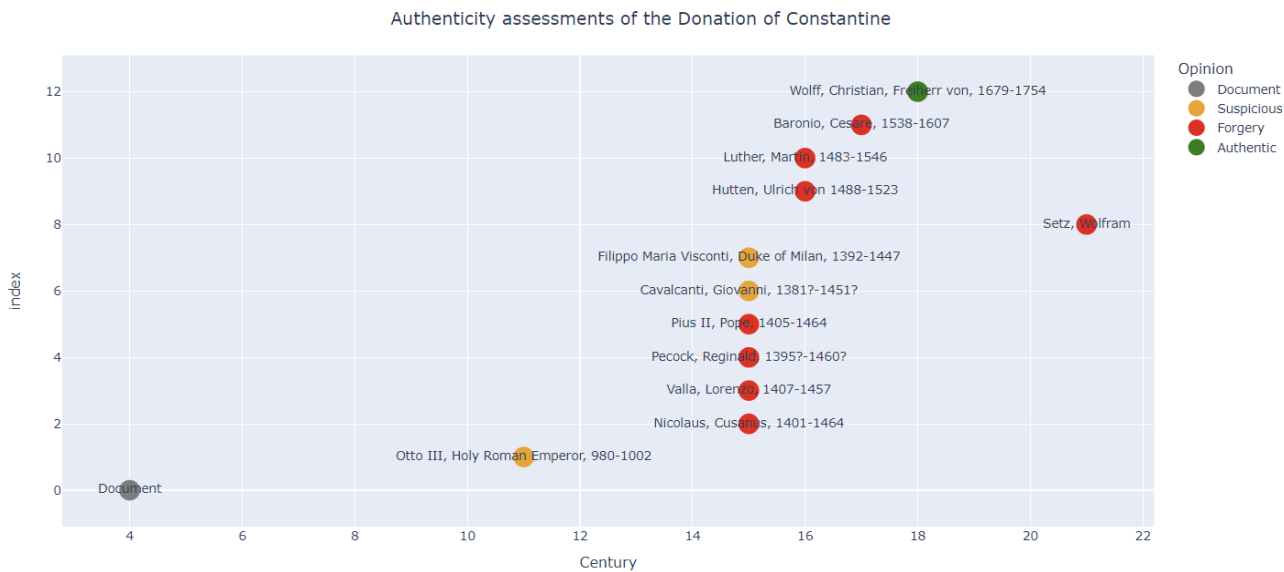

Figure 2: Scatter plot representing claims about the *Donation of Constantine*

## 6. CONCLUSION AND FUTURE WORK

This study explored the generation of structured data about documents' authenticity assessment from natural language texts. In particular, the pipeline of this work exploits LLM to select, extract and classify relevant claims, and Semantic Web technologies to structure and validate them. The LLMs offer a simple interface for some of the more complex NLP-related tasks (such as claim extraction and classification) that are otherwise impossible, given the lack of training data for this case study. They are valuable tools inside a pipeline, but using them as an End-to-End model. Errors happen, and avoiding them is impossible without human-in-the-loop approaches. Fine-tuning might be a better solution with a slightly bigger dataset. NLP techniques (e.g. NER) and Semantic Web technologies (e.g. external alignments) are necessary to clean and homogenise the LLM outputs to create a fully structured and interrogable dataset. In conclusion, the *forgont* ontology and the resulting KG demonstrate that highly structured data on the topic of documents' authenticity assessment allows for further queries and analysis on the topic, as shown in Figure 2. This document authenticity assessment serves as a case study highlighting how structuring a critical debate opens a set of new challenges in knowledge extraction. Furthermore, it explores new possibilities in computation, e.g. via SPARQL queries, aiming to understand how scholarly opinions have shaped contemporary knowledge.

## ACKNOWLEDGEMENTS

This project has been partially funded under the National Recovery and Resilience Plan (NRRP), specifically under Investment I.4.1 - Borse PNRR Patrimonio Culturale. The funding is from the Call for tender No. 351 of 9 April 2022 by the Italian Ministry of Culture, supported by the European Union – NextGenerationEU initiative.

We extend our gratitude to the Università degli Studi di Bologna for their administrative and academic support, particularly in facilitating the course "Patrimonio Culturale nell'Ecosistema Digitale" (Cultural Heritage in the Digital Ecosystem), Code: DOT22KM3LN, under the decree Ministeriale n. 351 del 9 aprile 2022.